\documentclass[preprint2]{aastex}

\slugcomment{}

\shorttitle{Jet and close environment of PKS 2155-304.}
\shortauthors{Liuzzo et al.}

\begin{document}


\title{On the radio and NIR jet of PKS 2155-304 and its close environment.}


\author{E. Liuzzo\altaffilmark{1}, R.Falomo\altaffilmark{2}, A. Treves\altaffilmark{3}, C. Arcidiacono\altaffilmark{4}, E. Torresi\altaffilmark{5}, M. Uslenghi\altaffilmark{6}, J. Farinato\altaffilmark{2}, A. Moretti\altaffilmark{2}, R. Ragazzoni\altaffilmark{2}, E. Diolaiti\altaffilmark{4}, M. Lombini\altaffilmark{4},  R. Brast\altaffilmark{7}, R. Donaldson\altaffilmark{8}, J. Kolb\altaffilmark{8}, E. Marchetti\altaffilmark{8} and S. Tordo\altaffilmark{8}}


\affil{$^{1}$ Istituto di Radioastronomia, INAF, via Gobetti 101, 40129 Bologna, Italy;\\$^{2}$ Osservatorio Astronomico di Padova, INAF, vicolo dell'Osservatorio 5, 35122 Padova, Italy;\\$^{3}$ Universit\`a dell'Insubria (Como), Italy, associated to INAF and INFN;\\$^{4}$ Osservatorio Astronomico di Bologna, INAF, Bologna, Via Ranzani 1, 40127 Bologna, Italy;\\$^{5}$ INAF/IASF Bologna, via Gobetti 101, 40129 Bologna, Italy;\\$^{6}$INAF/IASF Milano, via E. Bassini 15, 20133 Milano, Italy;\\$^{7}$ Dipartimento di Astronomia, Universit\`a di Bologna, via Ranzani 1, 40127 Bologna, Italy;\\$^{8}$ European Southern Observatory, Karl-Schwarschild-Str 2, 85748 Garching bei M\''{u}nchen, Germany.}

\email{liuzzo@ira.inaf.it}

\begin{abstract}
PKS 2155-304 is one of the brightest BL Lac object in the sky and a very well studied target from radio to TeV bands. We report on high resolution ($\sim$0.12 arcsec) direct imaging of the field of PKS 2155-304 using adaptive optics near-IR observations in J and Ks bands obtained with the ESO multi-conjugate adaptive optic demonstrator (MAD) at the Very Large Telescope. These data are complemented with archival VLA images at various frequencies to investigate the properties of the close environment of the source. We characterized the faint galaxies that form the poor group associated to the target. No radio emission is present for these galaxies, while an old radio jet at $\sim$20 kpc from the nucleus of PKS 2155-304 and a jet-like structure of $\sim$2 kpc ($\sim$1 arcsec) in the eastern direction are revealed. No counterparts of these radio jets are found in the NIR or in archival Chandra observations.
\end{abstract}


\keywords{galaxies: BL Lacertae objects: individual: PKS 2155-304 - instrumentation: adaptive optics}



\section{Introduction}\label{intro}

\begin{figure*}[ht!]
\epsscale{2}
\plotone{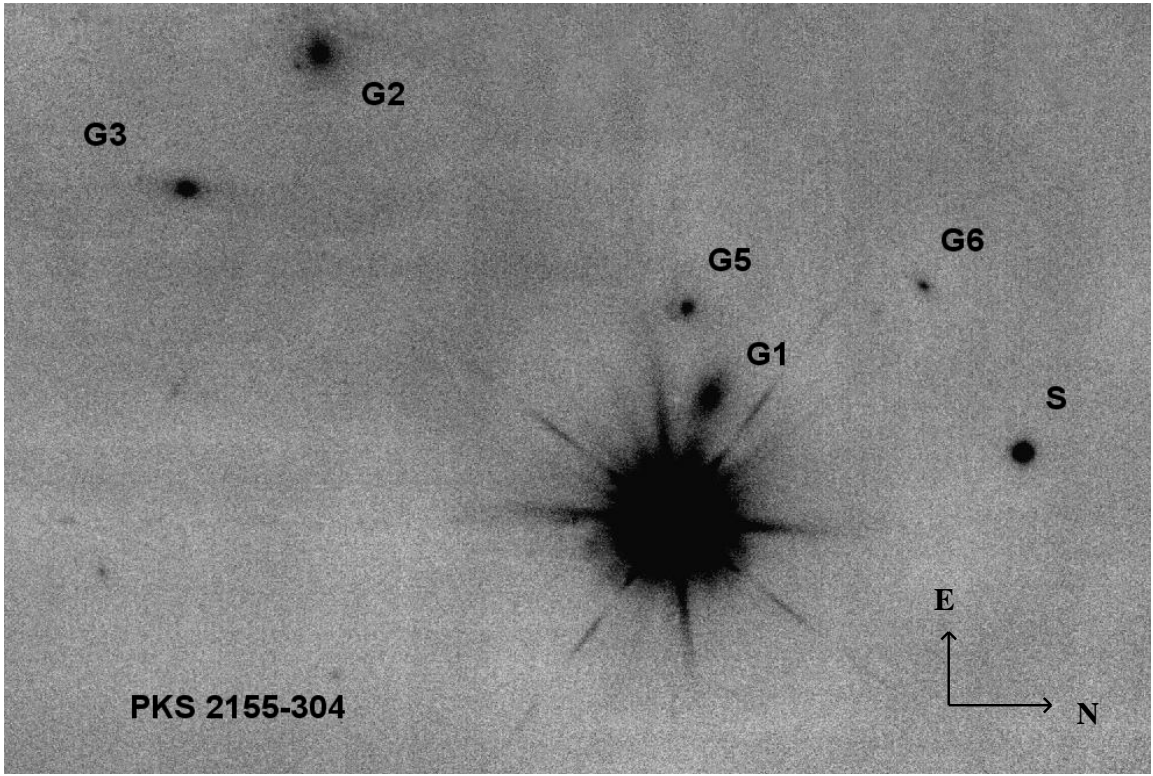} 
\caption{MAD image in Ks band of the field centered on PKS 2155-304 (the brightest object in the field). The figure covers $\sim$30$\times$40 arcsec$^{2}$ central portion of the observed field. Five galaxies are revealed in the field of the BL Lac. We name some of them according to Falomo et al. 1993 (see comments in Section 2).} \label{fig1}
\end{figure*}

\begin{figure}[ht!]
\epsscale{1}
\plotone{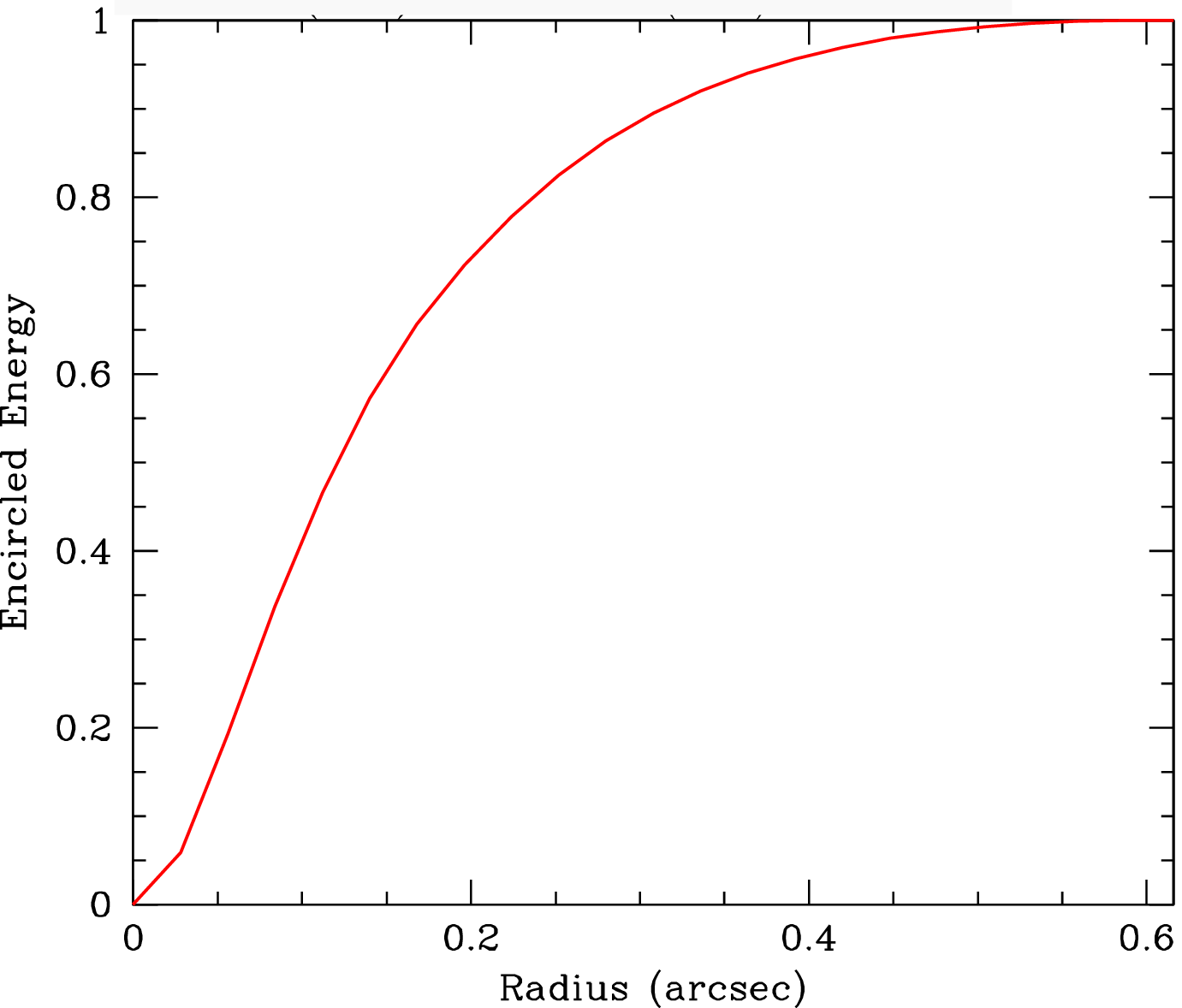} 
\caption{Encircled energy distribution obtained for the only star in the Ks MAD image of PKS 2155-304 (see Figure 1). 50\% of encircled energy is gathered within a radius of 0.13 arcsec.}\label{encS}
\end{figure}

PKS 2155-304 (V$\sim$13, z=0.116) is one of the best studied BL Lac objects from radio to TeV bands (Urry et al. 1993, 1997; Aharonian 2005, 2007, 2009; Foschini et al. 2007, 2008; and reference therein). It is an archetypal X-ray-selected BL Lac object well known to be rapidly and strongly variable throughout the whole electromagnetic spectrum on diverse timescales (e.g. Aharonian et al. 2005; Marshall et al. 2001;  Piner et al. 2004, 2008, 2010, and references therein). Many researchers have also found evidence of ultra-relativistic outflows (Aharonian et al. 2007, Foschini et al. 2007, Sakamoto et al. 2008).

At radio frequencies, the low resolution ($\sim$2 arcsec) Very Large Array (VLA) images presented by Ulvestad et al. (1986) and Laurent-Muehleisen et al. (1993) show an extended ($\sim$1 arcmin) halo around the core. At higher resolution ($\sim$0.6 arcsec), VLA images reveal a knot at 10 arcsec in the NW direction nearly 180$^{\circ}$ misaligned from the parsec-scale jet reported by Piner et al. (2010) (and references therein).  At millisarcsecond scale, Very Long Baseline Interferometer (VLBI) maps are presented by several authors (Ojha et al. 2004, 2010; Piner et al. 2004, 2008, 2010). Apparent speeds of jet components are also measured (e.g. Piner et al. 2010), and jet bending on parsec scales is found. 

In the optical and near-IR bands, PKS 2155-304 was observed with various instruments to study its host galaxy (Falomo et al. 1991, Kotinailen et al. 1998). It was found that the object is hosted by a luminous elliptical galaxy (M$_{R}$ = -24.4 and M$_{H}$= -26.8) at a redshift of 0.116 (Falomo et al. 1993, Sbarufatti et al 2006). The field around the object is moderately rich of galaxies. Spectroscopy of some of them show that they are at the same redshift of the BL Lac source indicating that its host galaxy is the dominant member of a group of galaxies. The closest companion galaxy is found 4.6 \arcsec ($\sim$9.7 kpc) from the source (Falomo et al. 1991, 1993).

Curiously while several BL Lac objects were imaged by Hubble Space Telescope (HST) cameras during various campaigns (see e.g. Scarpa et al 
2000), PKS 2155-304 was never observed with the HST and the best available high resolution images in the optical band remain those obtained by Falomo et al. (1991) with 0.7 arcsec of resolution.  

With the aim of investigating the close environment of PKS 2155-304, we obtained high resolution ($\sim$0.12 arcsec) NIR images with the innovative Multi Conjugate Adaptive Optics Demonstrator (MAD) system at ESO (European Southern Observatory) VLT (Very Large Telescope). In particular, we search for a NIR counterpart of the radio jet as found with a similar technique in a few other nearby BL Lac objects (PKS 0521-365 and PKS 2201+044, Falomo et al. 2009, and Liuzzo et al. 2011a, Liuzzo et al. 2011b). The only other object of this class showing an optical jet is 3C 371 (Scarpa et al. 1999, Sambruna et al.2007). To complement the new NIR images we also perform a detailed analysis of archival VLA images and compare with the new NIR data.

In the following, we adopt the concordance cosmology with $\Omega_m$=0.3, $\Omega_\lambda$=0.7 and H$_{0}$= 70 km s
$^{-1}$ Mpc$^{-1}$. In this scenario, at z=0.116, 1 arcsec corresponds to 2.1 kpc.

\section{MAD near-IR data.} \label{IR}

We performed J and Ks band observations (see {\bf Figure} 1) of PKS 2155-304
on 2007 September 28 using the ESO MAD mounted at UT3 (Melipal) of the VLT. MAD is a prototype for Multi
Conjugate Adaptive optics (MCAO) system which aims to
demonstrate the feasibility of different MCAO reconstruction techniques in the framework of the European Extremely Large Telescope (E-ELT) concept and the 2nd Generation VLT Instruments. MAD is designed to
perform wide Field of View (FoV) adaptive optics correction in J, H and Ks band over 2 arcmin on the sky by using
relatively bright (m(V)$<$ 14-16) Natural Guide Stars (NGS).
We refer to Marchetti et al. (2003) for a detailed description of MAD. The MCAO correction was obtained using Layer Oriented Multi-Pyramid Wave Front Sensor for
the MCAO reconstruction (Ragazzoni 1996,
Ragazzoni et al. 2000, Marchetti et al. 2005, Arcidiacono et al. 2008). The
detector used is an Hawaii-II and it has a 57 \arcsec $\times$ 57 \arcsec  FoV with pixel size of 0.028 \arcsec.

\begin{figure*}[ht!]
\epsscale{2}
\plotone{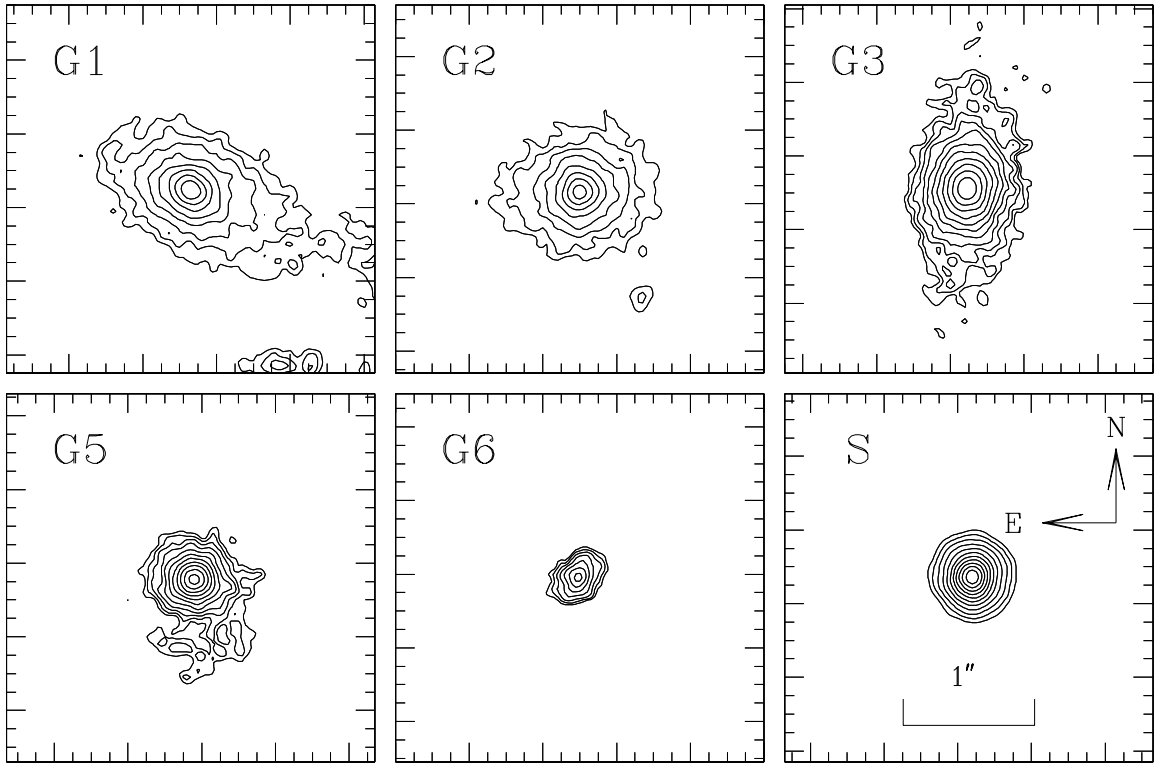} \label{fig3}
\caption{Contours images for the 5 galaxies plus the only star observed in the field of PKS 2155-304 in our Ks MAD observations.} 
\end{figure*}

\begin{table*}[ht!]
\begin{center}
\caption{Fundamental parameters of five galaxies in the field of PKS 2155-304.} \label{tab_nir1}
\begin{tabular}{cccccrc}
\tableline\tableline
Galaxy     & RA(J2000)& DEC(J2000)& z & Angular Dist. &Notes  \\
&&&& (arcsec)&\\
\tableline
G1 & 21 58 52.385 & -30 13 30.52  & 0.117             & 4.6 & Falomo et al. 1993\\
G2 & 21 58 53.334 & -30 13 44.65  & -                 & 21.0  & Falomo et al. 1993\\
G3 & 21 58 52.958 & -30 13 49.47  & 0.116             & 21.3 & Falomo et al. 1993 \\
G5 & 21 58 52.633 & -30 13 31.36  &  -                & 7.6 & This work\\
G6 & 21 58 52.697 & -30 13 22.83  &   -               & 12.3& This work\\         
\tableline\tableline 
\end{tabular}
\tablenotetext{}{Col. 5: Angular distance in arcsec from PKS 2155-304.}
\end{center}
\end{table*}

For these new observations, MAD was configured in the classical single reference approach (also called SCAO-Single Conjugated Adaptive
Optics) since the target is very bright, dominated by the unresolved nucleus, and in the AO FoV there are no other sufficiently bright stars. This is a different method than that used for previous MAD observations of extragalactic sources (Falomo et al. 2009, Liuzzo et al. 2011a, Liuzzo et al. 2011b). The pyramid wave-front sensor measured an R magnitude 13 once the adaptive optics
loop was closed. The pyramid wave-front
sensor measurements were correcting the first 35 Zernike modes using the pupil
conjugated 60 actuators Deformable Mirror. PKS 2155-304 was centered on the Infrared detector, and a five-position (dice 5)
dithering pattern with 5 arcsec step was applied in order to allow a reliable sky subtraction.
The dithering pattern was repeated 6 times both for the J and Ks filters, and
changing three times the starting central position of dice 5, in order to reduce
the effect of detector artifacts improving sky background evaluation.
The total integration time in each band was 1800 seconds. The Differential Image Motion Monitor (DIMM) measured 0.85 arcsec seeing at the zenith in the V band. We
obtained 3$\%$ and 12$\%$ Strehl Ratio values in the J and Ks bands, respectively corresponding to images with a core of 0.15 and
0.12 arcsec full width at half maximum (FWHM). In Figure 2, we show the encircled energy distribution derived from the only star in the field (see Figure 1). Half of the total energy is found within R(50$\%$)$\sim$0.13 arcsec. Due to the faintness of this star, no information about the shape of the Point Spread Function (PSF) is available at radii larger than 0.8 arcsec. This prevents performing an appropriate modeling of the emission around the target to characterize the host galaxy which extends up to $\sim$5 arcsec from the nucleus (Falomo et al. 1991, Kotilanein et al. 1998).

\begin{table}[h!]
\caption{Basic data for the five galaxies in the field of PKS 2155-304.} \label{tab_nir2}
\label{tab:radio}
\begin{center}
\begin{tabular}{ccccc}
\tableline\tableline
Galaxy    & Ks& J&  R$_{eff}$ & e     \\
          &   &  & (arcsec)     & \\
\tableline
G1 & 16.1& 16.9 & 0.78 & 0.5 \\
G2 & 16.1& 17.0 & 0.65 & 0.1 \\
G3 & 16.9& 18.1 & 0.30 & 0.5 \\
G5 & 17.7& 18.5 & 0.65 & 0.2 \\
G6 & 19.5& 19.4 & 0.62 & 0.6 \\         
\tableline\tableline 
\end{tabular} 
\end{center}
\scriptsize
\tablenotetext{}{Magnitudes are obtained with a precision of $\sim$0.1 mag.}
\tablenotetext{}{Col. 4: R$_{eff}$ is the effective radius from Ks band;}
\tablenotetext{}{Col. 5: e indicates source ellipticity.}
\end{table}

Since J and Ks band images are similar, we show here only images in the Ks band (Figures 1 and 3).
A few close-by galaxies are present in the observed field (Figures 1 and 3). Some of them were already detected by Falomo et al. (1991, 1993) and we use their labels for these galaxies (see Table 1). Another companion galaxy (G4) is located at $\sim$2 arcmin at south of PKS 2155-304 (out of the FoV of our MAD images). This is a bright galaxy (R=16.5) at the same redshift of the BL Lac object (Falomo et al. 1993). J and Ks band properties of the five galaxies are reported in Table 2. G1 and G3 are at the same redshift of PKS 2155-304 (see Table 1). For the other galaxies no spectroscopy is available. Assuming these galaxies are at the same distance of the BL Lac object and a standard H-K color, they could represent low luminosity (absolute magnitude in H band M$_{H}$ in the range [-22.4;-19.0] vs M$_{H}$= -26.8 for PKS 2155-304, Kotilainen et al. 1998) and small (R$_{eff}$ $\sim$1.5 kpc) satellite galaxies.

\section{Radio data} \label{radio}

To characterize the arcsec morphology of PKS 2155-304, we retrieved from National Radio Astronomy Observatory (NRAO) archive VLA radio observations at different frequencies with resolution comparable to that of our NIR MAD images. We analyzed simultaneous data obtained in L (1.4 GHz), C (5 GHz), X (8.4 GHz), U (15 GHz) and K (22.5 GHz) bands with VLA BnA configuration on 1994 May 14 (Project ID AK0359), and one observation in X band with VLA A configuration taken on 2003 July 24 (Project ID AR0517).

The target was observed for $\sim$6.5 minutes at 1.4 and 4.8 GHz, while the
8.4, 15 and 22.5 GHz observations have a total time on source of $\sim$40 minutes.
To produce our final images, we calibrated the data with the standard procedures in Astronomical Information Processing System (AIPS), accurately edited the visibilities,
and carried out a few self– calibration cycles.
The final images are produced with the IMAGR task. Image parameters are given in Table \ref{tab_radio}. Typical errors for flux measurements are $\sim$3$\%$ for frequencies lower than 8.4 GHz and $\sim$10$\%$ for the higher ones. In Figures 4-5, we present the most significant (L, C, and X bands) among our analyzed multifrequencies VLA images. In particular, maps in Figure 4 and in the top panel of Figure 5 are taken from 1994 data, while the X band high-resolution (Figure 5, bottom panel) is obtained from 2003 observations. Comparing the total flux densities at the two epochs (1994 and 2003, Table 3), PKS 2155-304 shows a clear evidence of variability in its non thermal continuum emission, which is a common feature for BL Lac objects (Urry \& Padovani 1995) and it is consistent with previous observations (e.g. Kedziora-Chudczer et al. 2001).

Concerning the morphology, at low resolution (L band, Figure 4  top panel), PKS 2155-304 is dominated by a bright core
with a peak flux density of 0.38 Jy/beam and a second component is present at 12 arcsec from the core in the NW direction with a flux density of $\sim$0.3 Jy. We also made a tapered image where an eastern diffuse emission is barely visible corresponding to the EW structure found by Ulvestad et al. 1986. A core plus the NW knot are also revealed in our C band data (Figure 4 bottom panel). 
At intermediate frequency (X band, $\sim$0.7 arcsec resolution), the NW component observed at lower frequencies is dimly visible only in tapered images of 1994 epoch (Figure 5, top panel), where the eastern jet is slightly resolved. However, in 2003 maps ($\sim$0.2 arcsec) the eastern jet clearly appears with an extension of $\sim$1 arcsec from the nucleus and a total flux density of $\sim$0.01 Jy (Figure 5, bottom panel).

At higher frequency (U band, $\sim$0.5 arcsec resolution), the jet component is resolved with an angular dimension of $\sim$2 arcsec, flux density of $\sim$0.01 Jy and it is oriented in the same direction of the mas structure found e.g. by Piner et al. (2010). 
Finally, in K band map ($\sim$0.2 arcsec resolution), the source is unresolved (peak flux density of $\sim$0.4 Jy/beam) and the jet is undetected up to 3.2 mJy/beam.

The eastern jet is also revealed in our analysis of archival Very Long Baseline Array (VLBA) L band data of 2000 January 15 with a final resolution of $\sim$ 14 mas, which is intermediate between VLBI and VLA published images up to now ($\sim$1 mas and $\sim$0.5 arcsec resolution, respectively). The structure is aligned with the sub-kpc jet (Figure 6 and Ulvestad
et al. 1983, Laurent-Muehleisen et al. 1993). From this alignment one could argue that black hole precession does not occur between these two angular scales.

In order to further investigate the nature of the source components, we made spectral index maps

\begin{table*}[th!]
\caption{Parameters of multifrequency radio images. } \label{tab_radio}
\label{tab:radio}
\begin{center}
\begin{tabular}{cllccc}
\tableline\tableline
data      & $\nu$  & HPBW                              & r.m.s.     & F$_{peak}$ & F$_{tot}$ \\
yy/mm/dd  &  GHz       & $\arcsec\times\arcsec,$$ ^{\circ}$ & mJy/b    &  Jy/b      & Jy        \\
\tableline
&&&&&\\
94/05/14  & 1.4  (L)      & 5.7 $\times$ 3.1, 44              &  0.3        & 0.38      &0.41          \\
94/05/14  & 4.8 (C)       & 3.0 $\times$ 3.0, 0               &  0.1        & 0.42      &0.44\\
94/05/14  & 8.4  (X)       & 1.1 $\times$ 0.7, 20              &  0.2        & 0.43      &0.44\\
94/05/14  & 15  (U)       & 0.5 $\times$ 0.4, 82              &  0.2        & 0.44      &0.45\\
94/05/14  & 22.5 (K)       & 0.2 $\times$ 0.2, 47              &  0.8        & 0.40      & 0.40\\
03/07/24  & 8.4 (X)       & 0.6 $\times$ 0.2, -22             &  0.1        & 0.21      & 0.22\\
&&&&&\\
\tableline\tableline 
\end{tabular}
\footnotesize
\tablenotetext{}{Col. 1: Observing data. Col. 2: Observing frequency.}
\tablenotetext{}{Col. 3: HPBW (half power beamwidth) and }
\tablenotetext{}{Position Angle of the major axis in deg.}
\tablenotetext{}{Col. 4: 1 rms level. Col. 5: Peak flux density.}
\tablenotetext{}{Col. 6: Total flux density.}
\end{center} 
\end{table*}

from the simultaneous multiwavelength data taken in 1994. For all maps that we produced, we used data closest in frequency and constructed with the same angular resolution.  
We define spectral index $\alpha$ as S $\sim$$\nu^{\alpha}$ where S is the flux density at frequency $\nu$. 
At arcsec scale, the NW knot shows a steep spectral index ($\sim$(-0.5)) that we interpreted as due to an old radio emission.
At sub-arcsec scale, the spectral indeces of the eastern jet vary radially from 0 to $-$1 according to the typical radiative and adiabatic losses of jets. 
At all resolutions, the central (0.5 arcsec) region has flat spectral index, while outside the $\alpha$ distribution is more complex with some asymmetries in the transverse direction suggesting the presence of large-scale helical magnetic field structure (c.f. Clausen$-$Brown et al. 2011).

\section{Comparison of radio and NIR images.}

We search for radio emission associated to the five close-by galaxies in the field of PKS 2155-304 (see Figure 1 and Table 1) using VLA archival data. No radio counterparts are found above the 1-mJy level.

In order to investigate the presence of NIR counterpart of the eastern radio jet, we show in Figure 6 the comparison between the MAD NIR image with the X band high resolution radio map (see also Figure 5). The MAD map is obtained subtracting a scaled point source modeled by the faint star north at $\sim$12.7 arcsec (see Figure 1). Due to the faintness of this star it is possible to define the PSF model only up to 0.8 arcsec from the center (Section 2 does not account for the prominent diffraction figure of the PSF.  At larger angular distance, we therefore used the extrapolation of model. In Figure 6 we show the complex structure of the PSF and the result of the subtraction of the model for the core of the PSF. 

In our MAD images, there is no counterpart in the NIR band of the eastern radio jet up to a magnitude limit of m$_{K}$= 16.5, while, in our previous studies of NIR counterparts of radio jets in BL Lac objects, we observed the brightest radio knot in PKS 0521-365 (Falomo et al. 2009) at distance from the core (d) $\sim$2 arcsec and m$_{K}$= 17.6, and in PKS 2201+044 (Liuzzo et al. 2011a) at d$\sim$2.2 arcsec and m$_{K}$= 18.5. 
Finally, we also investigated the morphology of PKS 2155-304 in the X-ray band. The Chandra X-ray satellite is the ideal instrument to this aim since it allows to perform spatially resolved studies thanks to its unprecedented angular resolution
(PSF FWHM 0.5 \arcsec). Fig. 7 shows a Chandra/HETG archival image with VLA C band contours map overimposed. We found that there are not evident structures possibly associated with the radio jet.

\acknowledgments

This work was supported by contributions of European Union, Valle D'Aosta Region and the Italian Minister for Work and Welfare.
This research has made use of the NASA/IPAC Extragalactic Data Base (NED), which is operated by the JPL, California Institute of Technology, under contract with the National Aeronautics and Space Administration.


\begin{figure}[ht!]
\epsscale{1.1}
\plotone{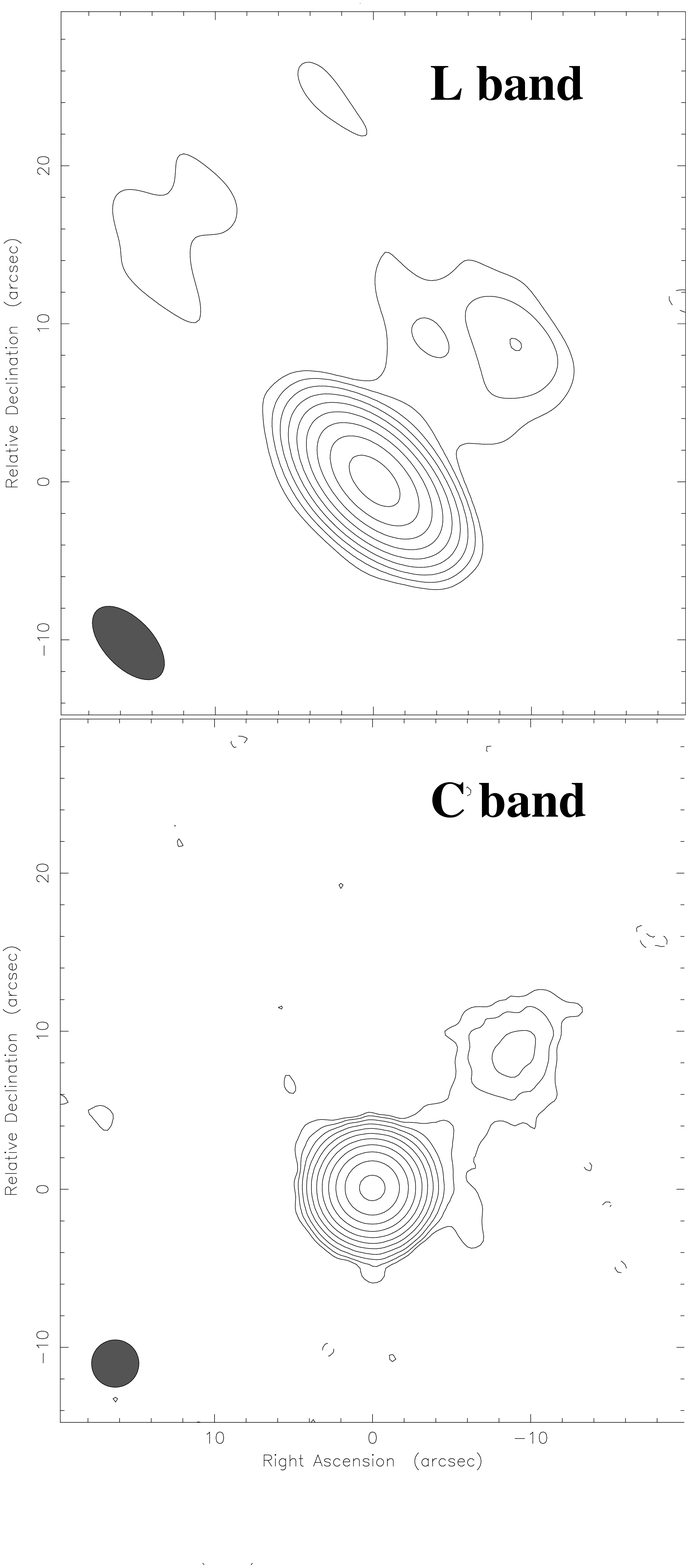} 
\caption{VLA radio contours images of PKS 2155-304 in L Band ({\it Top Panel}), C band ({\it Bottom Panel}). Noise levels, peak and total flux densities are reported in Table 3. Contours levels are:  1 $\times$ (1.1, 2.2, 4, 4.4, 8.8, 17.6, 35.2, 70.4, 140.8, and 281.6) mJy/beam for L Band; 0.4 $\times$ (1, 2, 4, 8, 16, 32, 64, 128, 258, 516, and 1032) mJy/beam for C band. North is to the top and East to the left.}\label{fig4}
\end{figure}

\clearpage 

\begin{figure*}[ht!]
\epsscale{1.1} 
\plotone{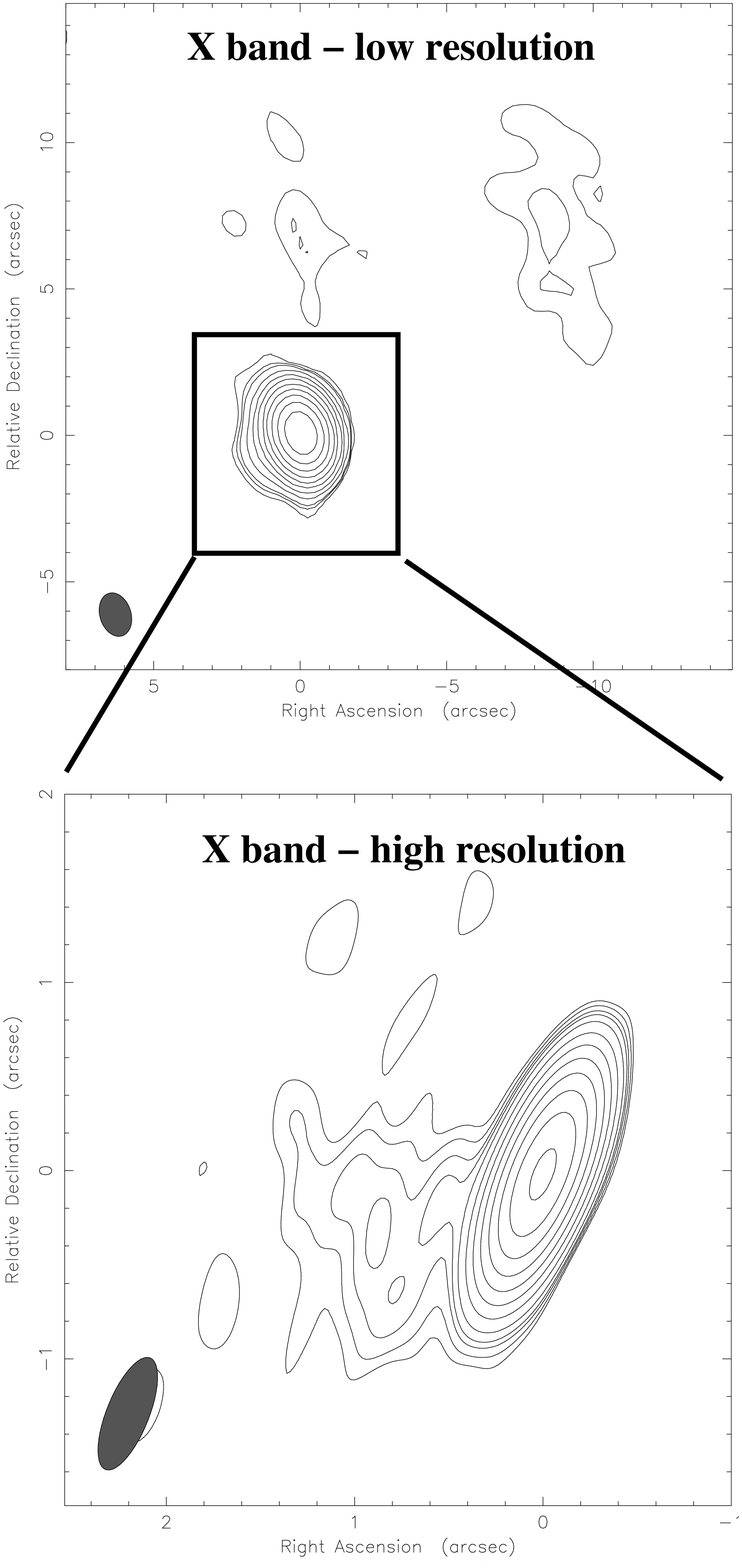} 
\caption{VLA radio contours images of PKS 2155-304 in X band low-resolution ({\it Top Panel}) and X band high-resolution ({\it Bottom Panel}) of the central region (rectangle in Top Panel). Noise levels, peak and total flux densities are reported in Table 3. Contours levels are:  0.38 $\times$(1.06, 1.6, 3.2, 6.4, 12.8, 25.6, 51.2, 102.4, 204.8, and 409.6 for X band low-resolution; 0.36 $\times$ (1.06, 1.6, 2.3, 2.8, 3.2, 6.4, 12.8, 25.6, 51.2, 102.4, 204.8, 409.6, and 819.2) mJy/beam for X band high-resolution.  North is to the top and East to the left.}\label{fig5}
\end{figure*}

\clearpage

\begin{figure*}[ht!]
\epsscale{1}
\plotone{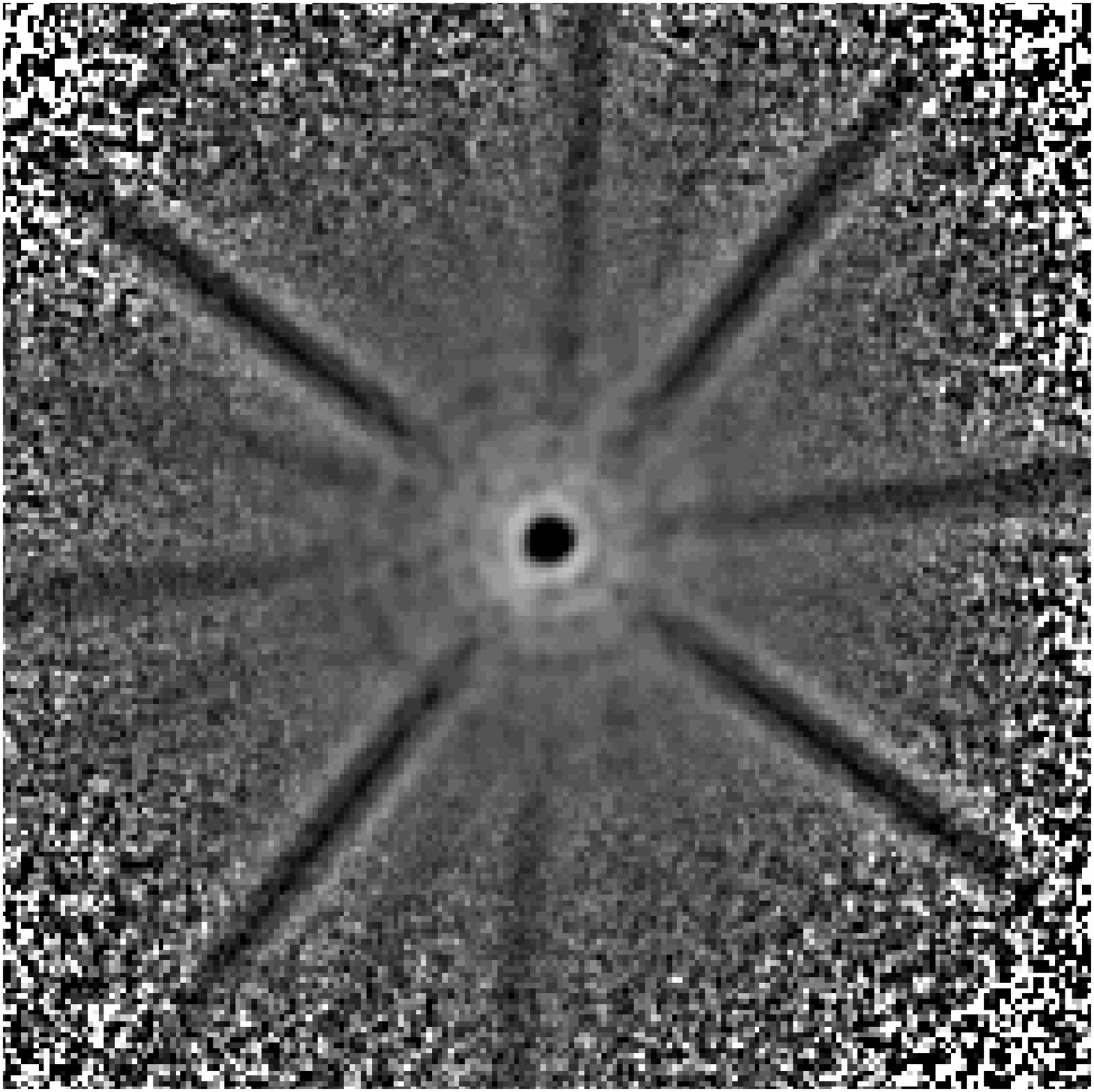}
\plotone{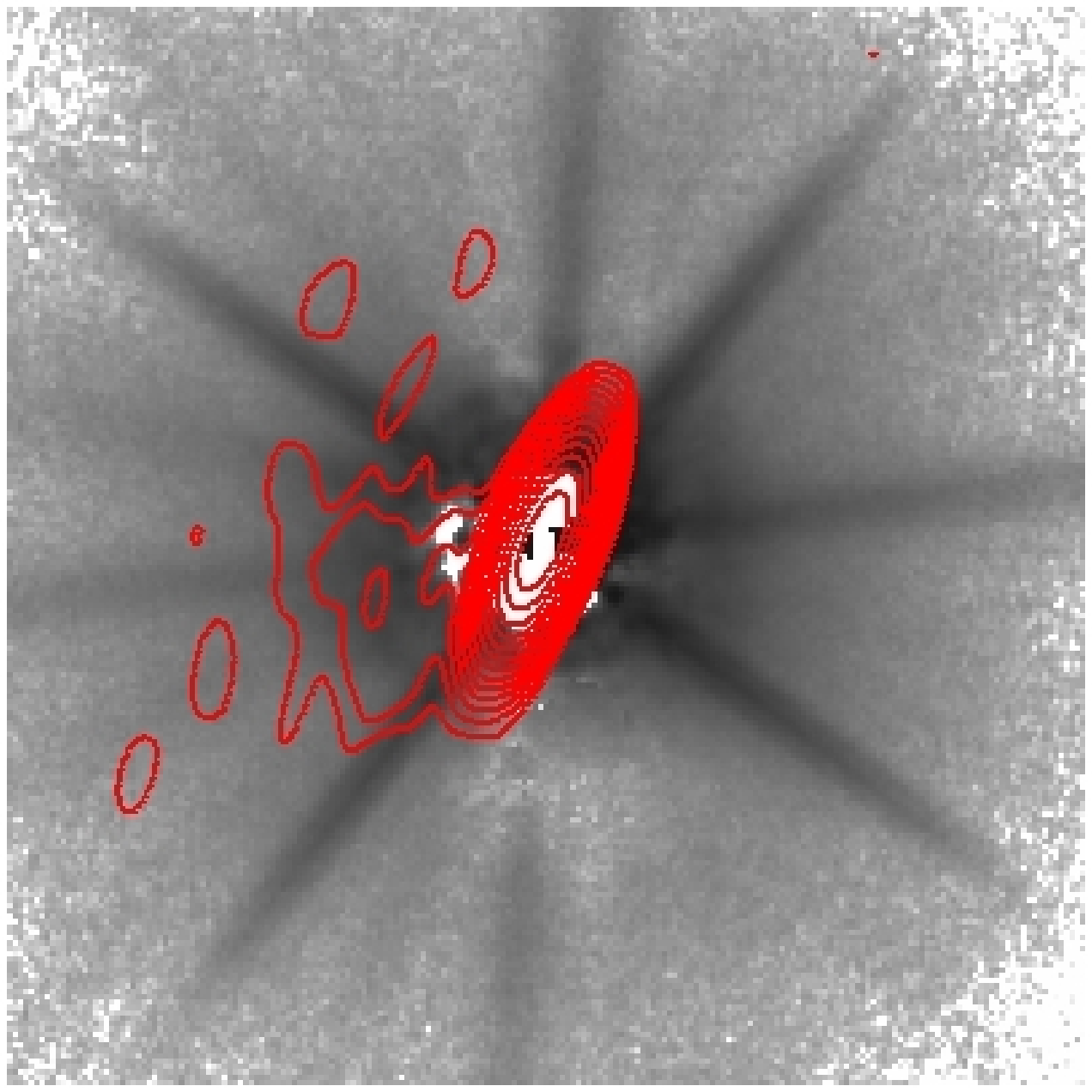} 
\caption{{\it Upper panel:} Image of PKS 2155-304 obtained using unsharp masking to enhance the high contrast features. The diffraction pattern due to the VLT secondary mirror and by the pupil mask of the Ground conjugated Deformable Mirror (the sharp one) are the most prominent components of the PSF. The inner region shows in the residual a rings pattern that follows both in ring intensity and radii the diffraction limited PSF profile in the Ks band. The spotty structure over the rings is common in adaptive optics corrected images: the diffraction patterns are modulated in intensity creating speckles.
{\it Bottom panel:} Contour image VLA map in X band low resolution of PKS 2155-304 overimposed to gray scale MAD image after the subtraction of a scaled PSF defined by the only star in the field of the Bl Lac object (see also Figure 1.)}\label{fig6}
\end{figure*}

\begin{figure*}[ht!]
\epsscale{1}
\plotone{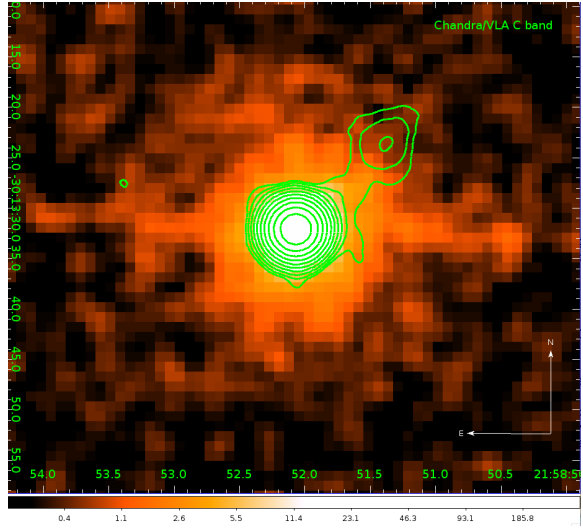}
\caption{Chandra/HETG image of PKS 2155-304 with VLA C band contours map overimposed. The X-ray observation was taken on 2008-07-02 for a total exposure of $\sim$15 ks. The X-ray image in the 0.4-8 keV range was extracted from the Transmission Grating and Catalog archive (TGCat
archive [http://tgcat.mit.edu/], Huenemoerder et al. 2011). The image was then smoothed with a Gaussian function of kernel radius 2.}\label{fig7}
\end{figure*}

\end{document}